\documentclass[11pt]{article}
\newcommand{\bm}[1]{\mbox{\boldmath $#1$}}
\renewcommand{\theequation}{\arabic{section}.\arabic{equation}}
 \def\bb{\bibitem} \def\lb{\label}
\def\be{\begin{equation}} \def\ee{\end{equation}}
\def\ba{\begin{eqnarray}} \def\ea{\end{eqnarray}} \def\part{\partial}
\def\nn{\nonumber}
\def\ol{\overline} \def\k{\kappa} \def\z{\zeta}
\def\L{\Lambda} \def\R{{\cal R}}
\def\X{{\bm X}} \def\J{{\bm J}} \def\bL{{\bm L}}
\def\a{{\bm \alpha}} \def\b{{\bm \beta}} \def\c{{\bm \gamma}}
\def\m{\ol{m}} \def\oL{\ol\L}

\begin{document}

\begin{titlepage}
\title{
\begin{flushright}\begin{small}    LAPTH-1312/09
\end{small} \end{flushright} \vspace{2cm}
Warped $AdS_3$ black holes \\ in new massive gravity}

\author{G\'erard Cl\'ement\thanks{Email: gclement@lapp.in2p3.fr}\\
\small {Laboratoire de  Physique Th\'eorique LAPTH (CNRS),} \\
\small {B.P.110, F-74941 Annecy-le-Vieux cedex, France}}
\date{24 March 2009}
\maketitle

\abstract{We investigate stationary, rotationally symmetric solutions
of a recently proposed three-dimensional theory of massive gravity.
Along with BTZ black holes, we also obtain warped $AdS_3$
black holes, and (for a critical value of the cosmological constant)
$AdS_2\times S^1$ as solutions. The entropy, mass and angular momentum
of these black holes are computed.}

\end{titlepage}\setcounter{page}{2}

\section{Introduction}

In a recent paper \cite{bht}, a new theory of massive gravity in three
dimensions has been proposed. In this theory, as in the case of the
older topologically massive gravity (TMG) \cite{djt}, the linearized
excitations about the Minkowski (or de Sitter or anti-de Sitter in the case
of a non-vanishing cosmological constant) vacuum describe a propagating
massive graviton. To the difference of TMG, which achieves this goal
through the addition to the Einstein-Hilbert action of a parity-violating
Chern-Simons term, new massive gravity (NMG) is a parity-preserving,
higher-derivative extension of three-dimensional general relativity. The
possibility of generalizing such an extension to higher dimensions has been
explored in \cite{no}, with the finding that only the three-dimensional
model is unitary in the tree level.

Cosmological TMG admits two very different kinds of black hole solutions,
BTZ black holes \cite{btz}, which are discrete quotients of $AdS_3$, and
warped $AdS_3$ black holes \cite{tmgbh,adtmg,ALPSS,tmgebh}, which are
discrete quotients of warped $AdS_3$. Similarly to BTZ black holes, these
have a four-parameter local isometry algebra, which generically is
$sl(2,R)\times R$, and may be generated from the corresponding
vacua by local coordinate transformations \cite{ALPSS,tmgebh}. The ADM
lapse function of these warped black holes goes to a constant value
at spacelike infinity (with the ADM shift function going to zero)
\cite{tmgbh}, which makes them closer in this respect to four-dimensional
black holes than the BTZ black holes. But the warped $AdS_3$ black holes
have the very special property of being intrinsically non-static, their
ergosphere extending to infinity \cite{tmgbh}.

It is straightforward to show that
cosmological NMG also admits the BTZ black holes as solutions \cite{bht}.
Because TMG and NMG have much in common, and indeed may be
unified in a ``general massive gravity'' model \cite{bht}, the existence of
warped $AdS_3$ black hole solutions to cosmological NMG may be
conjectured. The purpose of the present paper is to contruct these black
hole solutions, and to compute their mass, angular momentum, and entropy.

In the next section we investigate NMG with two Killing vectors,
write down in compact form the dimensionally reduced field
equations, and exhibit four constants of the motion. A simple ansatz
reduces in the third section these fourth order derivative equations
to a system of algebraic equations with two generic solutions,
leading either to BTZ black holes or warped $AdS_3$ black holes. In
the fourth section, we compute the mass, angular momentum and
entropy of these black holes, which satisfy the first law of black
hole thermodynamics, as well as a Smarr-like relation. Our results
are summarized in the Conclusion. Solutions with a horizon and
without naked closed timelike curves, but which may be transformed
to the vacuum by a global coordinate transformation and thus are not
genuine black holes, are discussed in the Appendix.

\section{New massive gravity with two Killing vectors}

The action of the cosmological new massive gravity theory is
\footnote{We choose the $(- + +)$ metric signature, so that some of
our signs differ from those of \cite{bht}.} \cite{bht} \be\lb{act}
I_3 = \frac1{16\pi G}\int d^3x \sqrt{|g|}\left[\R - \frac1{m^2}K -
2\L \right] \,, \ee where $\R$ is the trace of the Ricci tensor
$\R_{\mu\nu}$, the quadratic curvature invariant $K$ may be defined
in terms of the Schouten tensor $S_{\mu\nu} = \R_{\mu\nu} -
(1/4)g_{\mu\nu}\R$ as \be K =
\epsilon_{\lambda\mu\nu}\epsilon_{\rho\sigma\tau}g^{\lambda\rho}
S^{\mu\sigma}S^{\nu\tau} = \R_{\mu\nu}\R^{\mu\nu} - \frac38\R^2 \ee
and, for sake of generality, we will consider both signs of the real
squared mass parameter $m^2$.

We search for stationary circularly symmetric solutions of this
theory using the dimensional reduction procedure of \cite{EL}. A
three-dimensional metric with two commuting Killing vectors $\partial_t$,
$\partial_{\varphi}$ may be parametrized as
\be \lb{par}
ds^2=\lambda_{ab}(\rho)\,dx^a dx^b + \zeta^{-2}(\rho)R^{-2}(\rho)
\,d\rho^2
\ee
($x^0 = t$, $x^1 = \varphi$), where $R^2 = - \det\lambda$, and the scale
factor $\zeta(\rho)$ allows for arbitrary reparametrizations of the radial
coordinate $\rho$. The local isomorphism between the $SL(2,R)$ group of
transformations in the 2-Killing vector space and the Lorentz group $SO(2,1)$
suggests the parametrization of the $2 \times 2$ matrix $\lambda$
\be \lambda = \left(
\begin{array}{cc}
T+X & Y \\
Y & T-X
\end{array}
\right), \ee such that $R^2 \equiv \X^2$ is the Minkowski
pseudo-norm of the ``vector'' $\X(\rho) = (T,\,X,\,Y)$, \be \X^2 =
\eta_{ij}\,X^iX^j = -T^2+X^2+Y^2 \,, \ee  We shall use the notations
$\X\cdot{\bm Y} = \eta_{ij}\,X^iX^j$ for the scalar product of two
vectors, and  \be ({\X} \wedge {\bm Y})^i =
\eta^{ij}\epsilon_{jkl}X^k Y^l \ee (with $\epsilon_{012} = +1$) for
their wedge product.

The Ricci tensor components are \cite{adtmg} \be\lb{ric} {\R^a}_b =
-\frac{\z}2{\bigg( (\z RR')'{\bf 1}+(\z\ell)'\bigg)^a}_b\,, \quad
{\R^2}_2 = -\z(\z RR')'+\frac12\z^2(\X'^2)\,, \ee where $'$ denotes
the derivative $d/d\rho$, and $\ell$ is the matrix \be\lb{vecmat}
\ell = \left(
\begin{array}
[c]{cc} -L^{Y} & -L^T+L^X\\ L^T+L^X & L^{Y}
\end{array}\right)
\ee
associated with the vector
\be
\bL \equiv \X\wedge\X'\,.
\ee
It follows that
\ba\lb{RK}
\R &=& \z^2\left[-2(RR')'+\frac12(\X'^2)\right] - 2\z\z'RR'\,,\nn\\
K &=& \z^4\left[\frac12(\bL'^2) - \frac14(RR')'(\X'^2)+\frac5{32}(\X'^2)^2
\right] \\
&& + \z^3\z'\left[(\bL\cdot\bL')-\frac14RR'(\X'^2)\right] +
\z^2\z'^2\frac12(\bL^2)\,, \nn
\ea
reducing the action (\ref{act}) to the one-dimensional form
\ba
I_1 &=& \int d\rho \left[A\z^3 + B \z^2\z' + C\z\z^{'2} + D\z + E\z' +
2\z^{-1}\L\right] \lb{act1}\\
&=& \int d\rho \left[\left(A-\frac13B'\right)\z^3 + C\z\z^{'2} +
\left(D-E'\right)\z + 2\z^{-1}\L\right] \lb{act2} \ea (up to a
surface term), where \ba A &=&
\frac1{m^2}\left[\frac12(\X\cdot\X'')^2 - \frac12(\X^2)(\X''^2) -
\frac14(\X\cdot\X'')(\X'^2) - \frac3{32}(\X'^2)^2\right]\,, \nn\\
B &=& \frac1{m^2}\left[(\X\cdot\X')(\X\cdot\X'') - (\X^2)(\X'\cdot\X'')
- \frac14(\X\cdot\X')(\X'^2)\right]
\ea
(the expression of $C$ shall not be needed in the following, and those of
$D$ and $E$ are obvious from (\ref{RK})).

Rather than reducing the three-dimensional equations of massive
gravity \cite{bht}, it is more convenient to derive the reduced
equations directly from the reduced action. Taking advantage of the
reparametrization invariance of (\ref{par}), we shall fix the gauge
$\z =$ constant after variation of $\z$. The first form (\ref{act1})
of the reduced action is most convenient for variation relative to
$\X$, which leads to the equations \ba\lb{eqX} &&
\X\wedge(\X\wedge\X'''') + \frac52\X\wedge(\X'\wedge\X''')
+ \frac32\X'\wedge(\X\wedge\X''') \nn\\
&& \quad + \frac94\X'\wedge(\X'\wedge\X'') - \frac12\X''\wedge(\X\wedge\X'') \\
&& \quad - \left[\frac18(\X'^2) + \frac{m^2}{\z^2}\right]\X'' = 0
\,. \nn \ea The wedge product $\X\wedge(\ref{eqX})$ can be first
integrated, leading to the constancy of the super-angular momentum
vector \ba\lb{J} \J &=&
-\frac{\z^2}{m^2}\bigg\{(\X^2)[\X\wedge\X'''-\X'\wedge\X'']
+ 2(\X\cdot\X')\X\wedge\X'' \nn\\
&& + \left[\frac18(\X'^2) - \frac52(\X\cdot\X'')\right]\X\wedge\X'
\bigg\} + \X\wedge\X' \ea associated with the $SL(2,R)$ invariance
of the reduced action (\ref{act1}) \cite{part}. The second form
(\ref{act2}) of the reduced action is more convenient for variation
relative to $\z$, leading (in the gauge $\z =$ constant) to the
Hamiltonian constraint \ba\lb{ham} H &\equiv&
(\X\wedge\X')\cdot(\X\wedge\X''') - \frac12(\X\wedge\X'')^2
+ \frac32(\X\wedge\X')\cdot(\X'\wedge\X'') \nn\\
&& + \frac1{32}(\X'^2)^2 + \frac{m^2}{2\z^2}(\X'^2) + \frac{2m^2\L}{\z^4}
= 0\,.
\ea
The combination $2\X\cdot(\ref{eqX}) - 3(\ref{ham})$ leads to the simpler
scalar equation
\ba\lb{scal}
&& \frac12(\X\cdot\X'')^2 - \frac12(\X^2)(\X''^2) - \frac14(\X'^2)(\X\cdot\X'')
- \frac3{32}(\X'^2)^2 \nn\\
&& - \frac{m^2}{\z^2}\left[\frac32(\X'^2) + 2(\X\cdot\X'')\right] -
\frac{6m^2\L}{\z^4} = 0\,, \ea which is equivalent to the trace of
the three-dimensional field equations \cite{bht} \be K +m^2(\R-6\L)
= 0\,. \ee

\setcounter{equation}{0}
\section{Regular black holes}

As in the case of topologically massive gravity \cite{vuo,exact,nutgur,part},
the equations (\ref{eqX}), (\ref{ham}) presumably admit a variety of exact
solutions. We shall only construct here the solutions generated from the
quadratic ansatz
\cite{part}
\be\lb{an} \X =\a\,\rho^2 +
\b\,\rho + \c\,, \ee
(with $\a$, $\b$ and $\c$ linearly independent constant vectors), which in
the case of TMG leads to warped $AdS_3$ black holes \cite{tmgebh}. Inserting
this ansatz into the vector
equation (\ref{eqX}), we find that the fourth and third order
components vanish provided the vectors $\a$ and $\b$ are constrained
by \be \a^2 = 0\,, \qquad (\a\cdot\b) = 0\,, \lb{null}\,. \ee It is
easy to show that the two constraints (\ref{null}) further imply
\be\lb{b} \a\wedge\b = b\a\,, \qquad \b^2 = b^2\,, \ee for some
real constant $b$. The lower order components of (\ref{eqX}) and the
scalar equation (\ref{scal}) are then satisfied provided \ba &&
\left[z + \frac{17}8b^2 - \m^2\right]\a = 0\,, \lb{solvec}\\
&& z^2 + \frac14 b^2z - \frac3{64}b^4 - \m^2\left[\frac34b^2 -
2z\right] - 3 \m^2\oL\ = 0\,, \lb{solscal}\ea where we have put \be
(\a\cdot\c) = -z\,, \ee and $\m^2 \equiv m^2/\z^2$, $\oL \equiv \L/\z^2$.

Equation (\ref{solvec}) has two solutions. The first is $\a = 0$
(implying $z=0$), so that our ansatz (\ref{an}) reduces to
\be\lb{BTZ} \X = \b\,\rho + \c\,, \ee which leads to the BTZ black
hole metric \ba ds^2 & = & (-2l^{-2}\rho + {\rm M}/2)\,dt^2 - {\rm
J}\,dt\,d\varphi + (2\rho + {\rm M}l^2/2)\,d\varphi^2 \nonumber \\
& & + [4l^{-2}\rho^2 - ({\rm M}^2l^2-{\rm J}^2)/4]^{-1}\,d\rho^2\,,
\ea for $\z = 1$, $b^2 = 4l^{-2}$ \cite{part,tmgbh}. The $AdS_3$
curvature parameter $l^{-2}$ is obtained by solving (\ref{solscal}),
\be\lb{ell} l^{-2} = 2m^2\left[-1 \pm \sqrt{1-\L/m^2}\right] \ee (compare
with Eq. (29) of \cite{bht}). For the upper sign, this parameter is
positive if $\L < 0$ (with the restriction $\L > m^2$ for $m^2 <
0$), and reduces as it should to $l^{-2} = - \L$ for $m^2 \to
\infty$. The other branch of the solution (lower sign) exists only
in the tachyonic case $m^2 < 0$ for $\L > m^2$.

The second solution of (\ref{solvec}) is \be\lb{solz} z = \m^2 -
\frac{17}8b^2\,. \ee Inserting this in (\ref{solscal}), we obtain
the equation \be \m^4 - 3b^2\m^2 + \frac{21}{16}b^4 - \m^2\oL\ =
0\,, \ee which (assuming $\L/m^2  \neq 1$) is solved by \be\lb{mzb}
\frac{m^2}{\z^2b^2} = \frac{6 \pm
\sqrt{3(5+7\L/m^2)}}{4(1-\L/m^2)}\,. \ee with $b^2 \neq 0$. At this
point, we can without loss of generality fix the scale $\z$ and the
spatial orientation\footnote{The sign of $b$ can be reversed by
reversing the angular orientation $\varphi \to - \varphi$.} so that
\be b = -1\,, \ee which enables us to make contact with the results
of \cite{tmgebh}. Squaring (\ref{an}), we obtain \be R^2 =
(1-2z)\rho^2 + 2(\b\cdot\c)\rho + \c^2 =
\beta^{2}(\rho^2-\rho_0^2)\,, \ee where we have set \be\lb{z} z =
(1-\beta^2)/2\,, \ee translated the radial coordinate so that
$(\b\cdot\c)=0$, and defined $\c^2 \equiv -\beta^2\rho_0^2$. As
shown in the Appendix, we can choose a rotating frame and a
length-time scale such that \ba\lb{abc} \a &=& (1/2,-1/2,0)\,,\qquad
\b = (\omega,-\omega,-1)\,, \nn\\ \c &=& (z+u, z-u, -2\omega z)
\quad (u=\beta^2\rho_0^2/4z + \omega^2z) \ea (other possible choices
either can be reduced to this by a global coordinate transformation,
or lead to non-black hole solutions). This choice leads to the
metric \ba ds^2 &=& -\beta^2\frac{\rho^2-\rho_0^2}{r^2}\,dt^2 +
r^2\bigg[d\varphi - \frac{\rho+(1-\beta^2)\omega}{r^2}\,dt\bigg]^2
\nonumber \\ &&\qquad +
\frac1{\beta^2\z^2}\frac{d\rho^2}{\rho^2-\rho_0^2}\,, \lb{bh} \ea
where \be\lb{r2} r^2 = \rho^2 +2\omega\rho + \omega^2\,(1-\beta^2) +
\frac{\beta^2\rho_0^2}{1-\beta^2}\,, \ee and the constants $\beta^2$
and $\z$ are given by \be\lb{beta2} \beta^2 = \frac{9 - 21\L/m^2 \mp
2\sqrt{3(5+7\L/m^2)}}{4(1-\L/m^2)}\,, \quad \z^{-2} =
\frac{21-4\beta^2}{8m^2}\,. \ee The solutions in the special cases
$\beta^2 = 1$ and $\beta^2 = 0$ are given in \cite{tmgebh} (where
$\mu_E$ should be replaced by $\z$).

The metric (\ref{bh}) represents a black hole of the warped $AdS_3$
type, with two horizons at $\rho = \pm \rho_0$, if $\beta^2>0$ and
$\rho_0^2 \ge 0$. As discussed in \cite{tmgebh}, naked closed timelike
curves (CTC) do not occur provided $\beta^2 < 1$ and $\omega >
-\rho_0/\sqrt{1-\beta^2}$. Including the limiting cases $\beta^2 =
1$ and $\beta^2 = 0$ (which are also causally regular in a certain
parameter range), the necessary condition for the existence of
causally regular warped $AdS_3$ black holes is therefore \be\lb{range} 0 \le
\beta^2 \le 1 \ee (implying, from the second equation (\ref{beta2}),
$m^2 > 0$). This corresponds to the upper sign in
(\ref{beta2}) and
\be
-\frac{35m^2}{289} \le \L \le \frac{m^2}{21}\,.
\ee
Special values are $\L = -35m^2/289$, corresponding to $\beta^2 = 1$,
$\L = 0$, corresponding to $\beta^2 = (9-2\sqrt{15})/4$, and $\L = m^2/21$,
corresponding to $\beta^2 = 0$.

The case $\L = m^2$, for which Eq. (\ref{mzb}) becomes undeterminate, deserves
a special investigation. This is carried out in the Appendix, with the result
that (in the framework of our quadratic ansatz) the only solution without
naked CTC is $AdS_2\times S^1$.

\setcounter{equation}{0}
\section{Entropy, mass and angular momentum}
In this section, we shall compute the various thermodynamical
observables for the black holes of the previous section, and check
that they obey the first law. We start with the computation of the
entropy, which is a straightforward application of Wald's general
formula \cite{wald93,jkm,wald94} \be S = 2\pi\oint_h dx\sqrt{\gamma}
\frac{\delta{\cal L}}
{\delta\R_{\mu\nu\rho\sigma}}\varepsilon_{\mu\nu}\varepsilon_{\rho\sigma}
\,, \ee where $h$ is the spatial section of the event horizon,
$\gamma$ is the determinant of the induced metric on $h$, $\cal L$
is the Lagrangian in (\ref{act}), and $\varepsilon_{\mu\nu}$ is the
binormal to $h$. For a stationary circularly metric of the form
(\ref{par}), this gives \ba S &=& 4\pi A_h\left(\frac{\delta{\cal
L}}
{\delta\R_{0202}}(g^{00}g^{22})^{-1}\right)_h\nn\\
&=& \frac{A_h}{4G}\left(1 -
\frac1{m^2}\left[(g^{00})^{-1}\R^{00} + g_{22}\R^{22} -
\frac34\R\right]_h\right) \,, \ea where $A_h = 2\pi r_h$ is the
horizon ``area". Computing the Ricci tensor components from
(\ref{ric}), we obtain, on the horizon $R^2 = 0$, \be\lb{entro} S =
\frac{A_h}{4G}\left(1 +
\frac1{2\m^2}\left[(\X\cdot\X'') - \frac14(\X'^2)\right]\right)\,.
\ee

For the BTZ black hole (\ref{BTZ}), this leads to \be\lb{Sbtz} S =
\frac{A_h}{4G}\left(1 - \frac1{2m^2l^2}\right)\,, \ee
which could also be obtained directly from the results of
\cite{sasoda,sasen,park}. For the warped $AdS_3$ black holes (\ref{bh}),
the result can be simplified with the aid of (\ref{solz}) to
\be\lb{Sbh} S =
\frac{A_h}{2G\m^2}\,.\ee
It is remarkable that, although their
curvature is not constant\footnote{But their curvature invariants are
constant and depend only on the parameter $\beta^2$ \cite{ALPSS,tmgebh}.}
as in the BTZ case, their entropy is simply
the Bekenstein-Hawking entropy renormalized by a factor $2/\m^2 = 16/(21
- 4 \beta^2)$, independently of the black hole parameters $\rho_0$ and
$\omega$.

The computation of the mass and angular momentum of these black
holes is straightforward in the BTZ case, using e.g. the
Abbott-Deser-Tekin (ADT) approach to the computation of the energy
of asymptotically $AdS$ solutions to higher curvature gravity
theories \cite{AD,DT02}. In the case of the warped $AdS_3$ black
holes, an extension of the ADT approach to the case of massive
gravity with non-constant curvature backgrounds, similar to that
carried out in \cite{adtmg} for topologically massive gravity, is
required. In the present work, we shall make the educated guess
that, as in the case of generic three-dimensional Einstein-scalar
field theories \cite{black} and TMG \cite{adtmg}, the mass and
angular momentum can be computed from \be\lb{MJ} M =
-\frac{\zeta}{8G}(\delta J^Y + \Delta)\,, \quad J = \frac{\zeta}
{8G}(\delta J^T-\delta J^X)\,,  \ee where $\delta\J$ is the
difference between the values of the super-angular momentum
(\ref{J}) for the black hole and for the background. The term
$\Delta$, which depends on the specific theory considered, is not
known in the case of new massive gravity. However we can use
(\ref{MJ}) to compute the angular momentum $J$ (which does not
require the knowledge of $\Delta$), and derive the mass $M$ by
integrating the first law of black hole thermodynamics \be\lb{first}
dM = T_{H}dS + \Omega_{h}dJ\,, \ee where the Hawking temperature and
the horizon angular velocity, computed from the metric in ADM form,
are \be T_{H} =\frac{1}{4\pi}\zeta r_h(N^2)'|_h\,, \quad \Omega_{h}=
-N^{\varphi}|_h\,. \ee

In the case of the BTZ black hole (\ref{BTZ}), we obtain
\be
\J = \left(1 - \frac1{2m^2l^2}\right)\bL\,,
\ee
leading (if we assume that, as in the case of TMG, $\Delta = 0$ for the BTZ
black hole solution of new massive gravity) to
\be\lb{MJbtz}
M = \left(1 - \frac1{2m^2l^2}\right)\frac{{\rm M}}{8G}\,, \quad
J = \left(1 - \frac1{2m^2l^2}\right)\frac{{\rm J}}{8G}\,,
\ee
The mass, angular momentum and entropy are renormalized by the same factor
$(1-1/2m^2l^2)$ \cite{park} so that, contrary to the case of TMG where the
modification of the mass, angular momentum and entropy from the case of
cosmological gravity is non trivial \cite{tmgbh,solo}), the first law
(\ref{first}) is trivially satisfied. The integral Smarr-like relation
\be\lb{smarr}
M = \frac12T_H S + \Omega_h J
\ee
satisfied by the usual BTZ black holes \cite{cai} and black hole solutions
to generic three-dimensional Einstein-scalar field theories \cite{black},
as well as by BTZ black holes in TMG \cite{park1}, is also satisfied by BTZ
black holes in new massive gravity.

In the case of the warped $AdS_3$ black hole (\ref{bh}), we find for the
super-angular momentum $\J$
\be\lb{Jgen}
\J = - \frac{2\beta^2}{\m^2}\left[\b\wedge\c + \rho_0^2\a\right]\,,
\ee
leading (if the spacetime (\ref{bh}) with $\rho_0 = \omega = 0$ is chosen as
background) to
\be\lb{Jbh}
J = \frac{\zeta\beta^2}{4G\m^2}\left[\omega^2(1-\beta^2) -
\frac{\rho_0^2}{1-\beta^2}\right]\,.
\ee
Inserting this into the first law (\ref{first}), together with the entropy
(\ref{Sbh}) and the temperature and horizon angular velocity
\be
T_H = \frac{\zeta\beta^2\rho_0}{A_h}\,, \quad \Omega_h =
\frac{2\pi\sqrt{1-\beta^2}}{A_h} \quad \left(A_h =
\frac{2\pi}{\sqrt{1-\beta^2}}[\rho_0 + \omega(1-\beta^2)]\right)\,,
\ee
we obtain
\be\lb{Mbh}
M = \frac{\zeta\beta^2(1-\beta^2)}{2G\m^2}\omega\,,
\ee
which, as in the case of TMG \cite{adtmg}, is twice the ``naive''
super-angular momentum value ((\ref{MJ}) with $\Delta = 0$).
The fact that the integrability conditions for (\ref{first}) are satisfied
is a non-trivial check of our formula (\ref{MJ}) for the angular momentum.
These values of the mass, angular momentum and entropy satisfy the
modified Smarr-like formula \cite{tmgebh}
\be\lb{smarr2}
M = T_H S + 2\Omega_h J
\ee
appropriate for warped $AdS_3$ black holes. Let us also note that the
values (\ref{Mbh}), (\ref{Jbh}) and (\ref{Sbh})
for $M$, $J$ and $S$ coincide with the corresponding values for the warped
black holes of gravitating Chern-Simons electrodynamics (Eq. (5.15) of
\cite{tmgebh} with $\lambda \equiv \mu_E/2\mu_G = 0$) renormalized by a
factor $2/\m^2$.

\section{Conclusion}

In this paper, we have shown that, besides the BTZ black holes, the
cosmological new massive gravity theory of \cite{bht} also admits
warped $AdS_3$ black hole solutions, causally regular in the range
(\ref{range}), and computed their entropy, mass
and angular momentum. An interesting side result of our analysis, discussed
at the end of the Appendix, is that for the critical value $\L = m^2 < 0$
for which the two branches of the BTZ black hole solution coincide, NMG
also admits $AdS_2\times S^1$ as a solution.

The fact that our values for the warped black hole mass and angular
momentum satisfy non-trivially both the first law of black hole
thermodynamics and the modified Smarr relation (\ref{smarr})
indicates that these are very likely to be correct. This should be
confirmed by a computation from first principles, via e.g. an
extension of the ADT approach to the case of new massive gravity
with non-constant curvature backgrounds, along the lines followed in
\cite{adtmg}.

Another, related, problem which we feel should be addressed is that of the
sign of the energy. The massive excitations of TMG \cite{djt} or of NMG
\cite{bht} linearized about the Minkowski vacuum have negative energy, unless
the gravitational coupling constant $G$ is chosen to be negative. In
cosmological TMG linearized around a constant curvature background, either the
massive gravitons or the BTZ black holes have negative energy, except for
a critical, ``chiral'' value of the Chern-Simons coupling constant
\cite{chiral}. Similarly, irrespective of the sign of the gravitational
coupling constant, the sign of the energy of massive excitations of NMG
linearized around an $AdS_3$ background has been found to be opposite to the
sign of the mass (\ref{MJbtz}) and entropy (\ref{Sbtz}) of the BTZ black holes,
except for the critical value  $m^2l^2 = 1/2$ \cite{liusun}. However for this
value the BTZ black hole mass and entropy and the energy of the massive modes
all vanish, and the theory seems to be trivial. On the other hand, the
entropy (\ref{Sbh}) of the warped $AdS_3$ blackholes of NMG is positive
definite for $G > 0$, and their mass (\ref{Mbh}) is non-negative if
$\omega \ge 0$. A difficult task which should be addressed is that of the
linearization of the theory around an appropriate, non-constant curvature
background, e.g. warped $AdS_3$ or the warped vacuum $\rho_0 = \omega = 0$,
and the determination of the sign of the energy of the corresponding massive
excitations.

\renewcommand{\theequation}{A.\arabic{equation}}
\setcounter{equation}{0}
\section*{Appendix: Non-black hole solutions}

A generic null vector $\a$ can be parametrized by
\be\lb{apar}
\a = (c, c\cos\alpha, c\sin\alpha)\,,
\ee
with $c$ real and $ 0 \le \alpha < 2\pi$. From (\ref{par}) and (\ref{an}),
\be\lb{Vinf}
g_{\varphi\varphi} \sim c(1 - \cos\alpha)\rho^2 \qquad (\rho \to \infty)\,.
\ee
This is non-negative (absence of CTC at infinity) provided either 1) $\alpha
\neq 0$ and $c > 0$, or 2) $\alpha = 0$.

In the first case ($\alpha \neq 0$), transition to a rotating frame $d\varphi
\to  d\varphi' = d\varphi - \Omega dt$ preserves (\ref{Vinf}), but transforms
$\alpha^Y$ to 0 for $\Omega = \cot(\alpha/2)$. The transformed null
vector $\a'$ is of the form (\ref{apar}), where $\alpha'= \pi$ and $c'$ can
be set to the value $1/2$ by a combined length and time rescaling
\cite{tmgebh}, leading to the parametrization (\ref{abc}).

In the second case ($\alpha=0$), the sign of $c$ remains arbitrary, while its
absolute value can again be set to $1/2$ by a similar combined rescaling. The
generic vector $\b$ satisfying (\ref{b}) with $b=-1$ is of the form $\b =
(\omega,\omega, 1)$, where $\omega$ can be transformed to 0 by transition to a
co-rotating frame (which preserves $\a$ and $\beta^Y=1$) with angular velocity
$\Omega = \omega$. In this frame the vectors $\a$, $\b$ and $\c$ are
\ba\lb{abc2}
\a &=& \epsilon(1/2,1/2,0)\,,\qquad \b = (0,0,1)\,, \nn\\
\c &=& \epsilon(u+z,u-z,0) \quad (u=\beta^2\rho_0^2/4z)\,,
\ea
with $\epsilon =$ sign($z$). This leads
to the metric
\ba
ds^2 &=& -\frac{\beta^2}{|1-\beta^2|}(\rho^2-\rho_0^2)dt^2
+ |1-\beta^2|\bigg[d\varphi + \frac{\rho}{|1-\beta^2|}\,dt\bigg]^2
\nonumber \\ &&\qquad
+ \frac1{\beta^2\z^2}\frac{d\rho^2}{\rho^2-\rho_0^2}\,, \lb{rbr}
\ea
for all positive $\beta^2 \neq 1$. This metric, similar to the rotating
Bertotti-Robinson
metric of \cite{conform}, can be obtained from the first form of the
warped $AdS_3$ black hole metric given in Eq. (3.16) of \cite{tmgebh}
(with $c = \epsilon$ and $\omega = 0$) by exchanging the time and angular
coordinates, $t \leftrightarrow -\varphi$. It follows that the metric
(\ref{rbr}) can be generated from the vacuum ((\ref{rbr}) with $\rho_0 = 0$)
by a global coordinate transformation (Eq. (6.12) of \cite{tmgebh} with
$t \leftrightarrow -\varphi$), and so is not a black hole. Indeed, for
$\rho_0^2 = -\gamma^2 < 0$, Eq. (\ref{rbr}) can be transformed to Eq. (3.3)
of \cite{ALPSS} by putting $\rho = \gamma\sinh\sigma$ and rescaling the $t$
and $\varphi$ coordinates, showing that this spacetime is spacelike warped
$AdS_3$.  If, disregarding this fact, we compute the entropy (from
(\ref{Sbh})), and the mass and angular momentum (from (\ref{MJ}), where
$\Delta = \delta J^Y$ is assumed), we obtain
\be
S = -\frac{16\pi^2\sqrt{|1-\beta^2|}}{\k^2\m^2}\,, \quad M = 0\,, \quad
J = -\frac{4\pi\z\beta^2|1-\beta^2|}{\k^2\m^2}\,,
\ee
which, being independent of the (spurious) parameter $\rho_0$, trivially
satisfy the first law (\ref{first}) and, together with the
the temperature and horizon angular velocity
\be
T_H = \frac{\zeta\beta^2\rho_0}{2\pi\sqrt{|1-\beta^2|}}\,, \quad \Omega_h =
=-\frac{\rho_0}{|1-\beta^2|}\,,
\ee
also satisfy the modified Smarr-like relation (\ref{smarr2}).

In the case $\beta^2 = 0$, the solution (\ref{rbr}) is replaced by Eq. (3.22)
of \cite{tmgebh} with $c=1$, $\omega = 0$ and $t \leftrightarrow -\varphi$,
\be
ds^2 = -2\nu\rho dt^2 + \bigg[d\varphi + (\rho+\nu)dt\bigg]^2 + \frac{d\rho^2}
{2\z^2\nu\rho}
\ee
($\nu > 0$). This can similarly be generated from the corresponding vacuum
metric by a global coordinate transformation (Eq. (6.16) of \cite{tmgebh} with
$t \leftrightarrow -\varphi$).

In the limit $\L/m^2 \to 1$, Eq. (\ref{mzb}) still makes sense if
either 1) $b$ is fixed and the lower sign is chosen, leading in the
gauge $b^2=1$ to a non-causally regular black hole with $\beta^2 =
35/8$, or 2) $b \to 0$. For $\L/m^2 = 1$ and $b^2 = 0$, Eqs.
(\ref{solvec}) and (\ref{solscal}) are both solved by \be z =
\m^2\,. \ee Note that $b^2 = 0$ in (\ref{b}) implies $\b \propto
\a$, so that the ansatz (\ref{an}) can be reduced by a translation
of $\rho$ to \be\lb{BR} \X = \a\rho^2 + \c\,. \ee This leads to \be
R^2 = -2\m^2(\rho^2-\rho_0^2)\,, \ee with $\rho_0^2 = (\c^2)/2\m^2$,
so that the metric (\ref{an}) has the correct signature only for
$m^2 < 0$. The constant vectors $\a$ and $\c$ can, without loss of
generality, be chosen in the form \be \a = (\pm1/2, -1/2, 0)\,,
\quad \c = (\pm(\m^2-u), \m^2+u, v) \ee ($2u = \rho_0^2 -
v^2/2\m^2$). In the case of the upper sign, $g_{\varphi\varphi} =
T-X = \rho^2-2u < \rho^2-\rho_0^2$, so the metric has closed
timelike curves outside the horizon $\rho=\rho_0$. In the case of
the lower sign, $g_{\varphi\varphi} = T-X$ and $g_{t\varphi} = Y$
are both constant, so the parameter $v$ can be set to 0 by a frame
rotation, leading to the Bertotti-Robinson-like metric (in the gauge
$\z = 1$) \be\lb{BR1} ds^2 = -(\rho^2-\rho_0^2)dt^2 - 2m^2d\varphi^2
- \frac{d\rho^2}{2m^2 (\rho^2-\rho_0^2)} \quad (m^2 < 0)\,, \ee
which, for all real $\rho_0^2$, is a parametrization of $AdS_2\times
S^1$. In this case, the contribution to the entropy from the second,
quadratic term in the right-hand side of (\ref{entro}) exactly compensates
the Bekenstein-Hawking entropy, leading to a vanishing net entropy.
Similarly, the mass and angular momentum computed from (\ref{Jgen}) and
(\ref{MJ}) (with the same assumption as above) also vanish,
\be
S = 0\,, \quad M = 0\,, \quad J = 0\,,
\ee
and both the first law and the modified Smarr formula are trivially satisfied.

\end{document}